\documentclass[graybox]{svmult}

\usepackage{mathptmx}       % selects Times Roman as basic font
\usepackage{helvet}         % selects Helvetica as sans-serif font
\usepackage{courier}        % selects Courier as typewriter font
\usepackage{type1cm}        % activate if the above 3 fonts are
\usepackage{makeidx}         % allows index generation
\usepackage{graphicx}        % standard LaTeX graphics tool
\usepackage{multicol}        % used for the two-column index
\usepackage[bottom]{footmisc}% places footnotes at page bottom

\makeindex             % used for the subject index
\def\sumk{\mathop{\sum}\limits}

\begin{document}

\author{T. Vashakmadze}

\title*{On the Theory and Practice of Thin Walled Structures
}
\author{T. Vashakmadze}
\institute{T. Vashakmadze \at I. Vekua Institute of Applied
Mathematics of Iv. Javakhishvili Tbilisi State University, 2
University street, 0186, Tbilisi, Georgia; Faculty of Exact and
Natural Sciences of I. Javakhishvili Tbilisi State University, 2
University street, 0186, Tbilisi, Georgia
\email{tamazvashakmadze@gmail.com}}

\maketitle

\abstract*{We consider the problem of satisfaction of boundary conditions when the  generalized  stress vector is given on the surfaces for elastic plates and shells. This problem was open also  both for refined theories in the wide sense and hierarchical type models. This one for hierarchical models was formulated by Vekua. In nonlinear cases the bending and compression-extension processes did not split and for this aim we cited von
K$\acute{a}$rm$\acute{a}$n type system without variety of ad hoc assumptions since, in the classical form of this system of DEs one of them represents  the condition of  compatibility but  it is not an equilibrium equation. This problem was open also both for refined theories in the wide sense and hierarchical type models. Thus, we created the mathematical theory of refined theories both in linear and nonlinear cases for anisotropic nonhomogeneous elastic plates and shells, approximately satisfying the corresponding system of partial differential equations and boundary conditions on the surfaces. The optimal and convenient refined theory might be chosen easily by selection of arbitrary parameters; preliminarily a few necessary experimental measurements have been made without using any simplifying hypotheses. The same problem is solved for hierarchical models too.}

\abstract{We consider the problem of satisfaction of boundary conditions when the  generalized  stress vector is given on the surfaces for elastic plates and shells. This problem was open also  both for refined theories in the wide sense and hierarchical type models. This one for hierarchical models was formulated by Vekua. In nonlinear cases the bending and compression-extension processes did not split and for this aim we cited von K$\acute{\textrm{a}}$rm$\acute{\textrm{a}}$n type system without variety of ad hoc assumptions since, in the classical form of this system of DEs one of them represents  the condition of compatibility but  it is not an equilibrium equation. Thus, we created the mathematical theory of refined theories both in linear and nonlinear cases for anisotropic nonhomogeneous elastic plates and shells, approximately satisfying the corresponding system of partial differential equations and boundary conditions on the surfaces. The optimal and convenient refined theory might be chosen easily by selection of arbitrary parameters; preliminarily a few necessary experimental measurements have been made without using any simplifying hypotheses. The same problem is solved for hierarchical models too.}

\vspace*{0.5cm}

$\;\;\;\;\;\;\;\;\;\;\;\;\;\;\;\;\;\;\;\;\;\;\;\;\;\;\;\;\;\;\;\;\;\;\;\;\;\;\;\;
\;\;\;\;\;\;\;\;\;\;\;\;\;\;\;\;\;\;\;\;\;\;\;\;\;\;\;\;\;\;\;\;\;\;\;\;\;\;\;\;
\;\;\;\;\;\;\;\;\;\;\;\;\;\;\;\; Elasticity\;cannot\;be\;linear!$

$\;\;\;\;\;\;\;\;\;\;\;\;\;\;\;\;\;\;\;\;\;\;\;\;\;\;\;\;\;\;\;\;\;\;\;\;\;\;\;\;
\;\;\;\;\;\;\;\;\;\;\;\;\;\;\;\;\;\;\;\;\;\;\;\;\;\;\;\;\;\;\;\;\;\;\;\;\;\;\;\;
\;\;\;\;\;\;\;\;\;\;\;\;\;\;\;\;\;\;\;\;\;\;\;\;\;\; Ph.\;Ciarlet,\;[1,\;p.\;286]$

\section{Introduction}
\label{sec:1}

Let us consider the equilibrium equations of the elastic body in the form [1, 2]:
\begin{equation}
\partial_j(\sigma_{ij}+\sigma_{kj}u_{i,k})=f_i,\;\;\;x\in\Omega_h=D(x,y)\times\left]h^-(x,y),\;h^+(x,y)\right[.
\end{equation}
The boundary conditions:
\begin{equation}
T_{i3}=\sigma_{i3}+\sigma_{j3}u_{i,j}=g_i^{\pm},\;\;\;x\in S^\pm=D\times\{h^\pm\},\;\;\;T_3=(T_{13},\;T_{23},\;T_{33})^T,
\end{equation}
\begin{equation}
l\left[\partial_1,\;\partial_2,\;\partial_3\right](x,u)=g,\;\;\;x\in S=\partial D\times\left]h^-,\;h^+\right[.
\end{equation}
The relation between the displacement vector $u=(u_1,\;u_2,\;u_3)$, the symmetrical strain $\varepsilon$ and stress  $\sigma$ tensors satisfy the Cauchy formulae and Hooke's law:
\begin{equation}
\varepsilon_{ij}=\frac{1}{2}(u_{i,j}+u_{j,i}+u_{i,k}u_{j,k}),\;\;\;\varepsilon=A\sigma,\;\;\;\sigma=B\varepsilon.
\end{equation}
Above and below we used the basic notations according to [i.e. 2, pp. xiv-xv]
which are same to usually notations from well-known books and articles. For example, the repetition of an the index denotes summation; small Latin and Greek indices assume the values of $1, 2, 3$ and $1, 2$ accordingly, unless
otherwise stipulated. In a reference to a subsection of a section the first number denotes the number
of the section, the second one denotes the subsection. $\displaystyle \frac{\partial}{\partial x_i}=\partial_i=_{,i}$  is a derivative by $x_i$,
$\displaystyle \frac{\partial}{\partial t}=\partial_t$ is the derivative with respect to time, $\delta_{ij}$ is Kronecker's symbol.
$A$, $B=A^{-1}$ named as the compliance and stiffness matrices, $\Omega_h(x)=D(x,y)\times]h^{-},h^{+}[$ is 3dim cylindrical domain,
$2h=h^{+}-h^{-}$ is a thickness, $S^\pm$, $S$ are face and lateral surfaces.

The paper is dedicated to the problem of the satisfaction of the boundary conditions on surfaces $S^{\pm}$ of the elastic plates. Although the main part of this problem was solved in [2], but some statements needs improvements. It is well known, that the problem of satisfaction of boundary conditions on surfaces is important for all refined theories of von K$\acute{\textmd{a}}$rm$\acute{\textmd{a}}$n-Mindlin-Reissner (KMR) type except Reissner's [3] and Ambartsumian's [4] models, which has evident gaps.   Wholly, this problem depends on the justification of the calculus of variations.  As it is known, the violation of Riemann "Dirichlet Principle (DP)" was shown in the considerable examples of Weierstass and Hadamard. In the case of the Dirichlet boundary condition the justification of DP was shown by Hilbert [5] (for the bilinear functionals) and Razmadze [6, 7] (for 1dim problems in general cases). In case of Neumann (natural) conditions the principle step was made by Rektorys [8]. Here we also study this problem and we constructed an example for the elastic plates when the stress vector is given on $S^{\pm}$ and found the exact solution. If we use the Legendre polynomials as a basis by means of Vekua method  [9] we obtain the unstable process. This fact demonstrates, the existence of "Vekua problem" with regard of the satisfaction of boundary conditions on $S^{\pm}$ and which was studied by Vekua carefully, but incompletely [9, ch. I, 11, ch. II, 2]. We have studied this problem for the hierarchical models in case of the isotropic homogeneous elastic plate. Also we investigated and defined the functional spaces of admissible solutions. The results are given below using some main statements from  [2, ch. II, 6.2-3].

\section{Investigation of Stability Problems of Vekua Models for Elastic Prismatic Shells}
\label{sec:2}

Over 20 years Vekua studied the problem of constructing of 2D hierarchical models for an arbitrary integer $N$,
especially for $N=0,1,2$  without using any physical and geometrical hypothesis. Different versions of
the models correspond  to  the linear theory of isotropic elastic plates and shells with the variable thickness according to (1)-(4) in monograph  [9]. Numerous of scientists has been worked on this problem (see i.e. references in [10]). Vekua used the following way: for the relations  (1), (3) by means of (2)  the Galerkin method was applied by using as the basis the system of the Legendre polynomials $\{p_n(x),\;p_n(\pm1)=(\pm1)^n\}$. In  addition, new expressions of the type (7.2 c), [9, ch. I, 7.2] were introduced. Those expressions were named as the normalized moments of the field of stresses which are coordinated with boundary conditions. By the expression (8.4 a,b), (8.9) [9, 8.1] which represent 2dim boundary value problems Vekua has constructed the approximate solutions of (1)-(4) in the form:
$$
\displaystyle (u,\sigma)=\sumk\limits^{\infty}_{s=0}\left(\mathop{u}\limits^{s}(x,y),
\mathop{\sigma}\limits^{s}(x,y)\right)p_s(z),
$$
which "are not compatible with boundary data on face surfaces $S^+,\;S^-$. Therefore these
approximations may prove to be rather rough values near the face surfaces" [9, page 79].
We called it as the "Vekua problem". In [9] the solution  corresponding to the hierarchical
BVPs for any integer $N$ by additional functions satisfying also the approximate system of
DEs was corrected. This function depends on the sum of differences of
Legendre polynomials with respect to indices in the form (11.7) [9, ch. 1]:
$$
U_0=A_m(x,y)(p_{m+1}(\zeta)-p_{m-1}(\zeta))+A_{m+1}(x,y)(p_{m+2}(\zeta)-p_{m}(\zeta)),
$$
$$
\displaystyle \zeta=\frac{z-\bar{h}}{2h},\;\;\;m>N+2.
$$
When $N$, $m$ tends to infinity, the problem is open.

You can look for another way of investigating this problem in [9, ch. II, 2].
Here for the displacement vector and stress tensor the Taylor series is
used near the point $z=0$ and the boundary conditions are approximately satisfied on the surfaces. Besides,
the case is considered when the approximation has the second order.

Let us consider the case when the boundary value problem of the theory of elasticity is a 1dim problem and thus we have: $\displaystyle u_1=u_2=\varepsilon_{\alpha i}=\sigma_{\alpha i}=f_\alpha=0$, $h=1$, $\sigma_{33}=(\lambda+2\mu)u_{3,3}$. Then we get the following boundary value problem:
\begin{equation}
-u''(x)=f(x),\;\;\;u'(-1)=\alpha,\;\;\;u'(1)=\beta.
\end{equation}
As $\displaystyle z(x)=u(x)-\frac{\alpha+\beta}{2}x-\frac{\beta-\alpha}{4}x^2+u_0$,
problem (5) is equivalent to the following one:
\begin{equation}
-z''(x)=f(x)+\frac{\beta-\alpha}{2},\;\;\;z'(-1)=z'(1)=0.
\end{equation}
For simplicity we assume that $\displaystyle f(x)-\frac{\beta-\alpha}{2}=p_1(x)$
and consider the following coordinate system:
$$
\begin{array}{l}
\displaystyle q_k(x)=-(2k+1)\int_{-1}^x(x-t)p_k(t)dt\\
\displaystyle=\frac{1}{2k+3}(p_{k+2}-p_k)-\frac{1}{2k-1}(p_k-p_{k-2}),\;\;\;k=0,1,2,...,
\end{array}{}
$$
$$
-q_0=\frac{1}{3}(p_2-p_0),\;\;\;-q_1=\frac{1}{5}(p_3-p_1),\;\;\;q'(\pm1)=0.
$$
We will find the solution of (6) as the set: $z(x)=\sumk\limits^{\infty}_{k=0}z_kq_k(x)$.
Then by the projective method
$$
\begin{array}{l}
\displaystyle (-z'',-q_0(x))=z'(x)q_0(x)|_{-1}^1=0,\\
\displaystyle (-z'',-q_1)=\int_{-1}^1z'(p_3'-p_1')dx=(p_1,p_3-p_1)\;\;\;\Rightarrow\\
\displaystyle -z_1+\frac{3}{7}z_3=-1,\\
(-z'',-q_2)=z'(x)q_2|_{-1}^1+\int_{-1}^1
\sumk\limits^{\infty}_{k=0}z_kq'_kp_1dx\;\Rightarrow\\
\displaystyle -\frac{1}{3}z_0+2\frac{3+7}{3\cdot7}z_2-\frac{1}{7}z_4=0,\\
\displaystyle -\frac{1}{4n-1}z_{2n-2}+2\frac{4n+1}{(4n-1)(4n+3)}z_{2n}-\frac{1}{4n+3}z_{2n+2}=0,\;\;(n=2,4,...),\\
\displaystyle z_1-z_3=-\frac{1}{3},\;\;-\frac{1}{5}z_1+\frac{14}{5+9}z_3-\frac{1}{9}z_5=\frac{1}{15},\\
\displaystyle -\frac{1}{4n-1}z_{2n-1}+\frac{2(4n+1)}{(4n-1)(4n+3)}z_{2n}-\frac{1}{4n+3}z_{2n+1}=0,\;\;\;(n=1,3,5,...)\;\Rightarrow\\
\displaystyle z_1=-\frac{1}{3},\;\;\;z_0=z_n=0,\;(n=2,3,...),
\end{array}{}
$$
as matrices of both systems are irresoluble and by the theorem of Olga Taussky-Todd are nonsingular ones.
Thus the solution of problem (6) has the following form:
$$
z(x)=\frac{1}{3}q_1(x),\;\textmd{i.e.}\;-z''(x)=p_1,\;z'(\pm1)=0.
$$
Now if in (5) we put $\displaystyle f(x)=p_1(x),\;\alpha=\beta$ and $\displaystyle u(x)=\sumk\limits^{\infty}_{k=0}u_kp_k(x)$
by using methodology of [9] we obtain $\displaystyle u(x)=\left(\frac{2}{5}+\alpha\right)p_1(x)-\frac{1}{15}p_3(x)$. The first summand presence here
demonstrates unstable process same to [8, ch. 21, example 21.2].

In [2, ch. II, 6.3] we investigated  the problem of construction and
justification of Vekua type systems  using methodology of [8] in case of natural conditions.

By using the Galerkin method  for DEs (1) we obtain that the components of the stress
vector $\sigma_3$ for systems of DEs considered in [9] and [2] are different. For models
from [9] the condition (2) is not satisfied as underlined in [9, 11].
Let us return to the initial problem (1)-(4) and consider the linear case.
In the above-mentioned works  was considered the case when the components
of the exterior tension vector $\sigma_3$ is given on $S^{\pm}$. The problems of satisfying
these boundary conditions for any approximations were different among proposed systems.
For some models they are natural, while  for others they appear to be the main ones in the sense of variational
methods (see Rektorys [8]). We construct a class of operator equations actually
coinciding with systems (7.9 a,b), (7.18 h,i) or (8.16) [9].
For the sake of brevity, we shall denote it by $(V)$.

Let us use this expansion into Fourier-Legendre for incomplete series components of
stress tensor. By virtue of boundary conditions on $S^{\pm}$ we have:
\begin{equation}
\displaystyle \sigma_{\alpha\beta}=\mathop{\sum}\limits^{\infty}_{k=0}
\mathop{\sigma_{\alpha\beta}}\limits^{\!\!\!\!\!\!\!s}p_s\left(\frac{z}{h}\right),
\end{equation}
\begin{equation}
\sigma_3=\frac{(h+z)g^++(h-z)g^-}{2h}
+\mathop{\sum}\limits^{\infty}_{s=1}\mathop{\sigma_{3j}}\limits^{\!\!\!\!s}
\left[p_{s+1}\left(\frac{z}{h}\right)-p_{s-1}\left(\frac{z}{h}\right)\right],
\end{equation}
At first we construct the basic Vekua type hierarchical 2-dim model
which approximates the linear boundary value problem for homogeneous
isotropic plates (for details see [2, Ch. II, part 6.3]). Then equilibrium
equations in terms of components of the stress tensor will be
equivalent to the following infinite system
\begin{equation}
\begin{array}{l}
\displaystyle c_mh\mathop{\sigma_{\alpha\beta,\beta}}\limits^{\!\!\!\!\!\!\!\!\!\!\!\!\!\!\!\!\!\!m}
+(2m+1)c_m\mathop{\sigma_{\alpha\beta}}\limits^{\!\!\!\!\!\!\!\!\!\!\!\!m}=\mathop{f_{\alpha}}\limits^{\!\!\!m}
-hc_0\delta_{m0}\frac{g^+_\alpha-g^-_\alpha}{2},\\
\displaystyle c_mh\left(\mathop{\sigma_{\alpha 3,\alpha}}\limits^{\!\!\!\!\!\!\!\!\!\!\!\!\!\!\!\!\!\!m-1}-
\mathop{\sigma_{\alpha 3,\alpha}}\limits^{\!\!\!\!\!\!\!\!\!\!\!\!\!\!\!\!\!\!m+1}\right)
+(2m+1)c_m\mathop{\sigma_{33}}\limits^{\!\!\!\!\!\!\!m}=\mathop{f_{3}}\limits^{\!m}
-hc_0\delta_{m0}\frac{g^+_{\alpha,\alpha}+g^-_{\alpha,\alpha}}{2}\\
\displaystyle -hc_1\delta_{m1}\frac{g^+_{\alpha,\alpha}-g^-_{\alpha,\alpha}}{2}-
hc_0\delta_{m0}\frac{g^+_{3}+g^-_{3}}{2}
\end{array}{}
\end{equation}
where
$$
\mathop{f}\limits^{m}=\int_{-h}^hf(x_1,x_2,t)p_m\left(\frac{t}{h}\right)dt,\;\;\;c_m=\frac{2}{2m+1},\;\;\;m=0,1,2,\cdot\cdot\cdot.
$$
Hooke's law takes the following form:
\begin{equation}
\begin{array}{l}
\displaystyle c_mh\mathop{\sigma_{11}}\limits^{\!\!\!\!\!m}=(\lambda+2\mu)c_mh\mathop{u_{1,1}}\limits^{\!\!\!\!\!\!\!\!\!\!\!m}+
\lambda c_mh\mathop{u_{2,2}}\limits^{\!\!\!\!\!\!\!\!\!\!\!m}+\lambda(2m+1)c_m\mathop{\sum}\limits_{k\geq m(2)}
\mathop{u_{3}}\limits^{\!\!\!\!\!\!k+1},\\
\displaystyle c_mh\mathop{\sigma_{12}}\limits^{\!\!\!\!\!m}=\mu hc_m\left(\mathop{u_{1,2}}\limits^{\!\!\!\!\!\!\!\!\!\!\!m}+
\mathop{u_{2,1}}\limits^{\!\!\!\!\!\!\!\!\!\!\!m}\right),\\
\displaystyle c_mh\mathop{\sigma_{22}}\limits^{\!\!\!\!\!m}=\lambda c_mh\mathop{u_{1,1}}\limits^{\!\!\!\!\!\!\!\!\!\!\!m}+
\displaystyle (\lambda+2\mu)c_mh\mathop{u_{2,2}}\limits^{\!\!\!\!\!\!\!\!\!\!\!m}+\lambda(2m+1)c_m\mathop{\sum}\limits_{k\geq m(2)}
\mathop{u_{3}}\limits^{\!\!\!\!\!\!k+1},\\
\displaystyle c_mh\left(\mathop{\sigma_{3\alpha}}\limits^{\!\!\!\!\!\!\!m-1}-
\mathop{\sigma_{3\alpha}}\limits^{\!\!\!\!\!\!\!m+1}\right)=\mu hc_m\mathop{u_{3,\alpha}}\limits^{\!\!\!\!\!\!\!\!\!\!\!m}+
\mu(2m+1)c_m\mathop{\sum}\limits_{k\geq m(2)}
\mathop{u_{\alpha}}\limits^{\!\!\!\!\!\!k+1}\\
\displaystyle -hc_0\delta_{m0}\frac{g^+_{\alpha}+g^-_{\alpha}}{2}-hc_1\delta_{m1}\frac{g^+_{\alpha}-g^-_{\alpha}}{2},
\end{array}{}
\end{equation}
$$
\begin{array}{l}
\displaystyle c_mh\left(\mathop{\sigma_{33}}\limits^{\!\!\!\!\!\!\!m-1}-
\mathop{\sigma_{33}}\limits^{\!\!\!\!\!\!\!m+1}\right)=\lambda hc_m\mathop{u_{\alpha,\alpha}}\limits^{\!\!\!\!\!\!\!\!\!\!\!m}+
(\lambda+2\mu)(2m+1)c_m\mathop{\sum}\limits_{k\geq m(2)}
\mathop{u_{3}}\limits^{\!\!\!\!\!\!k+1}\\
\displaystyle -hc_0\delta_{m0}\frac{g^+_{3}+g^-_{3}}{2}-hc_1\delta_{m1}\frac{g^+_{3}-g^-_{3}}{2}.\\
\end{array}{}
$$
Here and (often) below the following note is used:
$$
\mathop{\sum}\limits_{k\geq i(s)}
\mathop{u}\limits^{k}=\mathop{u}\limits^{i}+\mathop{u}\limits^{i+s}+\mathop{u}\limits^{i+2s}+\cdot\cdot\cdot,\;\;\;
\mathop{\sum}\limits_{k\leq i(s)}
\mathop{u}\limits^{k}=\mathop{u}\limits^{i}+\mathop{u}\limits^{i-s}+\mathop{u}\limits^{i-2s}+\cdot\cdot\cdot.
$$
Formulae (9) and (10) make it possible to obtain an explicit form of Vekua type system in
displacement components. For this purpose we use Hooke's law for values $\sigma_{3i}$ and condition (2).
We shall have:
$$
g^+_\alpha=\mu\sumk\limits^{\infty}_{k=0}\left(\mathop{u_{3,\alpha}}\limits^{\!\!\!\!\!\!\!\!\!\!\!\!\!k}+
\frac{k(k+1)}{2h}\mathop{u_{\alpha}}\limits^{\!\!\!\!\!\!k}\right),
$$
$$
g^-_\alpha=\mu\sumk\limits^{\infty}_{k=0}(-1)^k\left(\mathop{u_{3,\alpha}}\limits^{\!\!\!\!\!\!\!\!\!\!\!\!\!k}-
\frac{k(k+1)}{2h}\mathop{u_{\alpha}}\limits^{\!\!\!\!\!\!k}\right),
$$
and
$$
g^+_3=\sumk\limits^{\infty}_{k=0}\left(\lambda\mathop{u_{\alpha,\alpha}}\limits^{\!\!\!\!\!\!\!\!\!\!\!\!\!k}+
(\lambda+2\mu)\frac{k(k+1)}{2h}\mathop{u_{3}}\limits^{\!\!\!\!\!\!k}\right),
$$
$$
g^-_3=\sumk\limits^{\infty}_{k=0}(-1)^k\left(\lambda\mathop{u_{\alpha,\alpha}}\limits^{\!\!\!\!\!\!\!\!\!\!\!\!\!k}-
(\lambda+2\mu)\frac{k(k+1)}{2h}\mathop{u_{3}}\limits^{\!\!\!\!\!\!k}\right).
$$
We define values $g^+\pm g^-$, entering (9). We shall have:
\begin{equation}
\begin{array}{l}
\displaystyle g^+_\alpha+g^-_\alpha=2\mu\sumk\limits^{\infty}_{k=0}\left(\mathop{u_{3,\alpha}}\limits^{\!\!\!\!\!\!\!\!\!\!\!\!\!2k}+
\frac{(k+1)(2k+1)}{h}\;\;\mathop{u_{\alpha}}\limits^{\!\!\!\!\!\!2k+1}\right),\\
\displaystyle g^+_\alpha-g^-_\alpha=2\mu\sumk\limits^{\infty}_{k=0}\left(\;\;\mathop{u_{3,\alpha}}\limits^{\!\!\!\!\!\!\!\!\!\!\!\!\!2k+1}+
\frac{k(2k+1)}{h}\;\;\mathop{u_{\alpha}}\limits^{\!\!\!\!\!\!2k}\right),\\
\displaystyle g^+_3+g^-_3=2\sumk\limits^{\infty}_{k=0}\left(\lambda\mathop{u_{\alpha,\alpha}}\limits^{\!\!\!\!\!\!\!\!\!\!\!\!\!2k}+
(\lambda+2\mu)\frac{(k+1)(2k+1)}{h}\;\;\mathop{u_{3}}\limits^{\!\!\!\!\!\!2k+1}\right),\\
\displaystyle g^+_3-g^-_3=2\sumk\limits^{\infty}_{k=0}\left(\lambda\;\;\mathop{u_{\alpha,\alpha}}\limits^{\!\!\!\!\!\!\!\!\!\!\!\!\!2k+1}+
(\lambda+2\mu)\frac{k(2k+1)}{h}\mathop{u_{3}}\limits^{\!\!\!\!\!\!2k}\right),
\end{array}{}
\end{equation}
From equations (10), summing up the three last formulae, for values $\mathop{\sigma_{3\alpha}}\limits^{\!\!\!\!\!\!\!m}$ we obtain:
$$
\begin{array}{l}
\displaystyle \mathop{\sum}\limits_{s\leq m(2)}\left(\mathop{\sigma_{3\alpha}}\limits^{\!\!\!\!\!s-2}-
\mathop{\sigma_{3\alpha}}\limits^{\!\!\!\!\!s}\right)=-\mathop{\sigma_{3\alpha}}\limits^{\!\!\!\!\!m}=
\mu\mathop{\sum}\limits_{s\leq m(2)}\mathop{u_{3,\alpha}}\limits^{\!\!\!\!\!\!\!\!\!\!\!s-1}+\frac{\mu}{h}\mathop{\sum}\limits_{s\leq m(2)}(2s-1)
\mathop{\sum}\limits_{k\geq s(2)}\mathop{u_\alpha}\limits^{k}\\
\displaystyle -\frac{1}{2}(g^+_\alpha+g^-_\alpha)\mathop{\sum}\limits_{s\leq m(2)}\delta_{s-1,0}
-\frac{1}{2}(g^+_\alpha-g^-_\alpha)\mathop{\sum}\limits_{s\leq m(2)}\delta_{s-1,1}.
\end{array}{}
$$
Similarly
$$
\begin{array}{l}
\displaystyle -\mathop{\sigma_{33}}\limits^{\!\!\!\!\!m}=
\mu\mathop{\sum}\limits_{s\leq m(2)}\mathop{u_{\alpha,\alpha}}\limits^{\!\!\!\!\!\!\!\!\!\!\!s-1}+\frac{\lambda+2\mu}{h}
\mathop{\sum}\limits_{s\leq m(2)}(2s-1)\mathop{\sum}\limits_{k\geq s(2)}\mathop{u_3}\limits^{k}\\
\displaystyle -\frac{1}{2}(g^+_3+g^-_3)\mathop{\sum}\limits_{s\leq m(2)}\delta_{s-1,0}
-\frac{1}{2}(g^+_3-g^-_3)\mathop{\sum}\limits_{s\leq m(2)}\delta_{s-1,1}.
\end{array}{}
$$
In these expressions $\mathop{\sigma_{3\alpha}}\limits^{\!\!\!\!\!\!\!\!\!\!-1}=\mathop{\sigma_{3\alpha}}\limits^{\!\!\!\!\!0}=0$ is assumed.

Now, by using formulae (11) from the latter representations after some computations, we get
$$
\mathop{\sigma_{3\alpha}}\limits^{\!\!\!\!\!\!\!\!m}=
\mu\mathop{\sum}\limits_{s\geq (m+1)(2)}\left[\mathop{u_{3,\alpha}}\limits^{\!\!\!\!\!\!\!\!\!\!\!\!\!s}+\frac{1}{2h}
((s+1)(s+2)-m(m+1))\;\mathop{u_\alpha}\limits^{\!\!\!\!\!s+1}\right],
$$
$$
\mathop{\sigma_{33}}\limits^{\!\!\!\!\!\!\!\!m}=
\mathop{\sum}\limits_{s\geq (m+1)(2)}\left[\lambda\mathop{u_{\alpha,\alpha}}\limits^{\!\!\!\!\!\!\!\!\!\!\!\!\!s}+\frac{1}{2h}
(\lambda+2\mu)((s+1)(s+2)-m(m+1))\;\mathop{u_3}\limits^{\!\!\!\!\!s+1}\right].
$$

Taking into account the last formulae, as well as (10), after
obvious simplifications with respect to components of the
displacement vector we obtain the following infinite system of Vekua's differential equations:
$$
\begin{array}{l}
\displaystyle l_2\mathop{u_{+}}\limits^{\!\!\!\!\!\!m}+(\lambda+\mu)h^{-1}(2m+1)\mathop{\sum}\limits_{k\geq m(2)}\textmd{grad}\;\;
\mathop{u_{3}}\limits^{\!\!\!\!\!\!\!k+1}+\mu h^{-2}\frac{2m+1}{2}\\
\displaystyle \times\mathop{\sum}\limits_{k\geq m(2)}[k(k+1)-m(m+1)]\mathop{u_{+}}\limits^{\!\!\!\!\!\!k}=
\frac{1}{c_mh}\left[\mathop{f_{+}}\limits^{\!\!\!\!\!\!m}-\frac{g_+^+-g_+^-}{2}\delta_{m0}\right],\\
\displaystyle \displaystyle \mu\Delta\mathop{u_{3}}\limits^{\!\!\!\!\!\!m}+(\lambda+\mu)h^{-1}(2m+1)\mathop{\sum}\limits_{k\geq m(2)}\textmd{div}\;\;
\mathop{u_{+}}\limits^{\!\!\!\!\!\!\!k}+(\lambda+2\mu) h^{-2}\frac{2m+1}{2}\\
\displaystyle \times\mathop{\sum}\limits_{k\geq m(2)}[k(k+1)-m(m+1)]\mathop{u_{3}}\limits^{\!\!\!\!\!\!k}=
\frac{1}{c_mh}\left[\mathop{f_{3}}\limits^{\!\!\!\!\!\!m}-\frac{g_3^+-g_3^-}{2}\delta_{m0}\right],
\end{array}{}
$$
Here
$$
u_+=(u_1,u_2)^T,\;\;f_+=(f_1,f_2)^T,\;\;g_+=(g_1,g_2)^T,
$$
$$
(l_2u_+,u_+)=\mu(\Delta u_\alpha,u_\alpha)+(\lambda+\mu)(\textmd{graddiv}u_+,u_+).
$$

From system (10), evidently, for values $\mathop{\sigma_{\alpha 3}}\limits^{\!\!\!\!\!\!\!\!m}$ we have:
$$
\begin{array}{l}
\displaystyle \mathop{\sigma_{\alpha 3}}\limits^{\!\!\!\!\!\!\!\!\!m-1}=\mathop{\sigma_{\alpha 3}}\limits^{\!\!\!\!\!\!\!\!\!m+1}
+\mu\mathop{u_{3,\alpha}}\limits^{\!\!\!\!\!\!\!\!\!\!\!\!m}+\mu\frac{2m+1}{h}\mathop{\sum}\limits_{k\geq m(2)}
\mathop{u_{\alpha}}\limits^{\!\!\!\!\!\!\!\!k+1}-\frac{1}{2}\delta_{m0}(g_\alpha^++g_\alpha^-)
-\frac{1}{2}\delta_{m1}(g_\alpha^+-g_\alpha^-)
\\
\displaystyle =\mathop{\sigma_{\alpha 3}}\limits^{\!\!\!\!\!\!\!\!\!m+3}
+\mu\mathop{u_{3,\alpha}}\limits^{\!\!\!\!\!\!\!\!\!\!\!\!m+2}+\mu\frac{2m+5}{h}\mathop{\sum}\limits_{k\geq m(2)}
\mathop{u_{\alpha}}\limits^{\!\!\!\!\!\!\!\!k+3}+\mu\mathop{u_{3,\alpha}}\limits^{\!\!\!\!\!\!\!\!\!\!\!\!m}+
\mu\frac{2m+1}{h}\mathop{\sum}\limits_{k\geq m(2)}
\mathop{u_{\alpha}}\limits^{\!\!\!\!\!\!\!\!k+1}\\
\displaystyle -\mathop{\sum}\limits_{k\geq m(2)}\left(\frac{g_\alpha^++g_\alpha^-}{2}\delta_{k0}+
\frac{g_\alpha^+-g_\alpha^-}{2}\delta_{k1}\right),\\
\displaystyle \mathop{\sigma_{\alpha 3}}\limits^{\!\!\!\!\!\!\!\!\!m}=\mathop{\sum}\limits_{k\geq m(2)}
\mathop{u_{3,\alpha}}\limits^{\!\!\!\!\!\!\!\!\!\!\!\!k+1}+\mu h^{-1}\mathop{\sum}\limits_{s\geq (m+1)(2)}
(2s+1)\mathop{\sum}\limits_{k\geq m(2)}\;\mathop{u_{\alpha}}\limits^{\!\!\!\!\!\!\!\!k+2}
\end{array}{}
$$
$$
\begin{array}{l}
\displaystyle -\frac{1}{2}\mathop{\sum}\limits_{k\geq (m+1)(2)}\left((g_\alpha^++g_\alpha^-)\delta_{k0}+
(g_\alpha^+-g_\alpha^-)\delta_{k1}\right),\;\;\;m=1,2,...\\
\displaystyle \mathop{\sigma_{\alpha 3}}\limits^{\!\!\!\!\!\!\!\!\!m}=\mu\mathop{\sum}\limits_{k\geq (m+1)(2)}
\left(\mathop{u_{3,\alpha}}\limits^{\!\!\!\!\!\!\!\!\!\!\!\!k}+\frac{1}{2}((k+1)(k+2)-m(m+1))\;
\mathop{u_{\alpha}}\limits^{\!\!\!\!\!\!\!k+1}\right).
\end{array}{}
$$
Analogously,
$$
\displaystyle \mathop{\sigma_{33}}\limits^{\!\!\!\!\!\!\!\!\!m}=\mathop{\sum}\limits_{k\geq (m+1)(2)}
\left(\lambda\mathop{u_{3,\alpha}}\limits^{\!\!\!\!\!\!\!\!\!\!\!\!k}+\frac{1}{2h}(\lambda+2\mu)((k+1)(k+2)-m(m+1))
\mathop{u_{3}}\limits^{\!\!\!\!\!k}\right).
$$
Taking into account these formulae we obtain
$$
\begin{array}{l}
\displaystyle \frac{1}{2}(g_\alpha^++g_\alpha^-)\mathop{\sum}\limits_{m=1}^{\infty}\delta_{m-1,0}+
\frac{1}{2}(g_\alpha^+-g_\alpha^-)\mathop{\sum}\limits_{m=1}^{\infty}\delta_{m-1,1}-
\mathop{\sigma_{3\alpha}}\limits^{\!\!\!\!\!\!\!\!\!m}
\\
\displaystyle =\mu\mathop{\sum}\limits_{k<(m+1)(2)}
\left(\mathop{u_{3,\alpha}}\limits^{\!\!\!\!\!\!\!\!\!\!\!\!k}+\frac{1}{2h}(k+1)(k+2)\;
\mathop{u_{\alpha}}\limits^{\!\!\!\!\!\!\!k+1}\right).
\end{array}{}
$$
Hence for values $\mathop{\sigma_{\alpha 3}}\limits^{\!\!\!\!\!\!\!\!\!m}$ we have:
$$
\begin{array}{l}
\displaystyle \mathop{\sigma_{\alpha 3}}\limits^{\!\!\!\!\!\!\!\!\!m}=
\frac{1}{2}(g_\alpha^++g_\alpha^-)\mathop{\sum}\limits_{m\geq1(1)}^{\infty}\delta_{m-1,0}+
\frac{1}{2}(g_\alpha^+-g_\alpha^-)\mathop{\sum}\limits_{m\geq1(1)}^{\infty}\delta_{m-1,1}
\\
\displaystyle -\mu\mathop{\sum}\limits_{k\leq(m-1)(2)}
\left(\mathop{u_{3,\alpha}}\limits^{\!\!\!\!\!\!\!\!\!\!\!\!k}+\frac{1}{2h}(k+1)(k+2)\;
\mathop{u_{\alpha}}\limits^{\!\!\!\!\!\!\!k+1}\right).
\end{array}{}
$$
Similarly for $\mathop{\sigma_{33}}\limits^{\!\!\!\!\!\!m}$ we shall have:
$$
\begin{array}{l}
\displaystyle \mathop{\sigma_{33}}\limits^{\!\!\!\!\!\!m}=
\frac{1}{2}(g_3^++g_3^-)\mathop{\sum}\limits_{m\geq1(1)}^{\infty}\delta_{m-1,0}+
\frac{1}{2}(g_3^+-g_3^-)\mathop{\sum}\limits_{m\geq1(1)}^{\infty}\delta_{m-1,1}
\\
\displaystyle -\mathop{\sum}\limits_{k\leq(m-1)(2)}
\left[\lambda\mathop{u_{\alpha,\alpha}}\limits^{\!\!\!\!\!\!\!\!\!\!\!\!k}+\frac{\lambda+2\mu}{2h}(k+1)(k+2)\;
\mathop{u_{3}}\limits^{\!\!\!\!\!\!\!k+1}\right].
\end{array}{}
$$
Taking into account these expression in (10) we obtain the
infinite system of differential equations according to Vekua's system $(V)$
in the following form:
\begin{equation}
\begin{array}{l}
\displaystyle l_2\mathop{u_{+}}\limits^{\!\!\!\!\!\!n}+h^{-1}(2n+1)\textmd{grad}\left(\lambda
\mathop{\sum}\limits_{i\geq n(2)}\mathop{u_{3}}\limits^{\!\!\!\!\!\!\!i+1}-\mu
\mathop{\sum}\limits_{i\leq n(2)}\mathop{u_{3}}\limits^{\!\!\!\!\!\!\!i+1}\right)\\
\displaystyle -\mu h^{-2}\frac{2n+1}{2}
\mathop{\sum}\limits_{i\leq n(2)}i(i+1)\mathop{u_{+}}\limits^{\!\!\!\!\!\!\!i}=\frac{1}{hc_n}\left[\mathop{f_{+}}\limits^{\!\!\!\!\!\!n}\right.\\
\displaystyle \left.-
\left(\frac{g_+^+-g_+^-}{2}\delta_{n0}+(g_+^++g_+^-)\mathop{\sum}\limits_{i\geq 1(1)}\delta_{i-1,0}+
(g_+^+-g_+^-)\mathop{\sum}\limits_{i\geq 1(1)}\delta_{i-1,1}\right)\right],\\
\displaystyle \mu\Delta\mathop{u_{3}}\limits^{\!\!\!\!\!\!n}+h^{-1}(2n+1)\textmd{div}\left(\mu
\mathop{\sum}\limits_{i\geq n(2)}\mathop{u_{+}}\limits^{\!\!\!\!\!\!\!i+1}-\lambda
\mathop{\sum}\limits_{i\leq n(2)}\mathop{u_{+}}\limits^{\!\!\!\!\!\!\!i-1}\right)\\
\end{array}{}
\end{equation}
$$
\begin{array}{l}
\displaystyle -(\lambda+2\mu)h^{-2}\frac{2n+1}{2}
\mathop{\sum}\limits_{i\leq n(2)}i(i+1)\mathop{u_{3}}\limits^{\!\!\!\!\!\!\!i}=\frac{1}{hc_n}\left[\mathop{f_{3}}\limits^{\!\!\!\!\!\!n}\right.\\
\displaystyle \left.-
\left(\frac{g_3^+-g_3^-}{2}\delta_{n0}+(g_3^++g_3^-)\mathop{\sum}\limits_{i\geq 1(1)}\delta_{i-1,0}+
(g_3^+-g_3^-)\mathop{\sum}\limits_{i\geq 1(1)}\delta_{i-1,1}\right)\right],
\end{array}{}
$$
$$
n=0,1,2,...,N.
$$

The comparison of these equations (12) with those of $(V)$ proves their identity
for $N=0,1,2$. When $N\geq3$, the main parts (containing only second order partial
derivatives)  of systems (7.18 h, i) [9] and (12) are different. Then [9, page 52] we
read: the (7.18 h, i) is a strong elliptic system of PDEs for $N\geq3$, "but we do not
rewrite this one in a more expanded form and shall not deal with the investigation
of problems of existence and uniqueness in the general form". Evidently, in order
to obtain effective values a priori in the form of energy inequalities for Vekua's
operator with fixed $N$ together with highest derivatives, we should pay attention to
the explicit form of summands with derivatives of zero and first order from unknown
moments $\mathop{u_{i}}\limits^{\!n}(x_1,x_2)\;(n=0,1,2,...)$ appearing in system (12). Thus, we constructed
(12) corresponding to the equations (1). Reduced boundary conditions, originated by
the data on the lateral surfaces $S$ and the construction of which is not difficult, should be
added to these systems. For this purpose we should multiply equalities (3) by
Legendre polynomials $\displaystyle p_i\left(\frac{z}{h}\right)$ and integrate them between $-h$ and $h$.
If Hooke's law and other representations from (4) are used, then we come up to the finite reduced
boundary conditions, defined on $\partial D$.

Now let us return to the linear case and use the variational principle, when the system
$(V)$ or (12) with reduced conditions (3) is the identity to the Euler-Lagrange equation
for the initial problem with the vector components of $\sigma_3$ given on surfaces $S^\pm$.

Let us consider the reduced systems, generated by the basic system (12). If we bound
ourselves $N+1$ vector equations, the following results will be true [2, ch. II, 6.3,
pp. 72-77].

\textbf{Theorem 1.} \textit{Let the boundary conditions on the lateral boundary $S$
corresponding to the linear problem (1)-(3), be homogeneous and  such that the equalities hold:}
\begin{equation}
\begin{array}{l}
\displaystyle \left(\mathop{u_{\alpha,\beta}}\limits^{\!\!\!\!\!\!\!\!\!\!\!\!\!n},\mathop{u_{i}}\limits^{\!\!\!m}\right)=
\int_D\mathop{u_{\alpha,\beta}}\limits^{\!\!\!\!\!\!\!\!\!\!\!\!\!n}\mathop{u_{i}}\limits^{\!\!\!m}dx_1dx_2=-
\left(\mathop{u_{\alpha}}\limits^{\!\!\!\!\!n},\mathop{u_{i,\beta}}\limits^{\!\!\!\!\!\!\!\!\!\!\!m}\right),
\end{array}{}
\end{equation}
\textit{where $\mathop{u_{i}}\limits^{\!\!\!n}$ are desired coefficients of the expansion function $u_i$.
Then the operator of the theory of plates, corresponding to the reduced systems (12),
satisfies an inequality of Korn's type with a constant, independent of $N$,}
\begin{equation}
\begin{array}{l}
\displaystyle -(L_NU_N,U_N)\geq\left(\left\|\textmd{grad}\mathop{U_{N}}\limits^{\!\!\!+}\right\|_1^2+
\left\|\textmd{div}\mathop{U_{N}}\limits^{\!\!\!+}\right\|_1^2+2\left\|\mathop{U_{N}}\limits^{\!\!\!3}\right\|_2^2\right),
\end{array}{}
\end{equation}
\textit{where}
\begin{equation}
\begin{array}{l}
\displaystyle \left(\mathop{u}\limits^{m},\mathop{v}\limits^{n}\right)_1=
\frac{1}{\sqrt{(2m+1)(2n+1)}}\left(\mathop{u}\limits^{m},\mathop{v}\limits^{n}\right),\\
\displaystyle \left(\mathop{u}\limits^{m},\mathop{v}\limits^{n}\right)_2=h^{-2}
\sqrt{(2m+1)(2n+1)}\left(\mathop{\sum}\limits_{i\geq m(2)}\mathop{u}\limits^{i+1},
\mathop{\sum}\limits_{i\geq n(2)}\mathop{v}\limits^{i+1}\right),\\
\displaystyle U_N=\left(\mathop{u}\limits^{0},...,
\mathop{u}\limits^{N}\right)^T=(u_1,u_2,u_3)^T=(u_+,u_3)^T,\;\;\;\mathop{u}\limits^{n}=0,\;n>N.
\end{array}{}
\end{equation}

\textbf{Theorem 2.} \textit{It is required to find the solution for the reduced system (12) in the domain $D(x_1,x_2)$,
satisfying homogeneous Dirichlet boundary conditions at $\partial D$ and being a trace of functions,
defined by inclusion $u_i(x,y,z)\in W_p^{2+\alpha}(\Omega_h),\;\alpha\geq0,\;p\geq1$.
Then is true the positive definiteness of Vekua's type operator for problem (1)-(3)
and will follow in case of Dirichlet's boundary conditions on $S$}
\begin{equation}
\begin{array}{l}
\displaystyle -(L_NU_N,U_N)\geq\mu\left(\kappa^2\left\|\mathop{U_{N}}\limits^{\!\!\!+}\right\|_1^2+
2\left\|\mathop{U_{N}}\limits^{\!\!\!3}\right\|_2^2\right),
\end{array}{}
\end{equation}
\textit{where $\kappa^2$ is a constant in Friedrich's inequality}

\textbf{Theorem 3.} \textit{It is required to find the solution for the reduced system (12) in $D$
 when the boundary of $\partial D$ domain is free and above same inclusion for $u_i$ is fulfilled.
Then we have}
$$
\begin{array}{l}
\displaystyle -\frac{1}{2h}(L_{\nu 1}U,U)\geq2\mu\left(\kappa_3^2\left\|U\right\|_1^2+
\kappa_2^2\left\|U\right\|_2^2\right),
\end{array}{}
$$
$$
\kappa_3^2=\kappa_1^2c_3,\;\;\;\kappa_1^2=\min\{c_1,c_2\},\;\;\;\kappa_2^2=\min\{1,c_2\},
$$
\textit{where $c_1,\;c_2$ are constants depending on $D$ and $c_3,\;c_4$ - the constants
of Poincar$\acute{e}$ inequality. We have used above the following notations:
$U=\left(\mathop{U}\limits^{+},\mathop{U}\limits^{3}\right)^T$, $L_{N}$, $L_{\nu 1}$ are operators,
corresponding to 2dim approximate systems (12), and 3dim linear problems (1), (3) for any $N\leq\infty$
when in (2): $\sigma_3^{\pm}=g^\pm=0$. }

Thus theorem 3 represents also a different proof of Korn's inequality.

In addition, for 1-dim models according (5) these system (8.16 b) ([9.8])
if $f\equiv0$ have the following form:
\begin{equation}
\begin{array}{l}
\displaystyle -(\lambda+2\mu)(2k+1)\left(\mathop{U'_{3}}\limits^{\!\!\!k+1}+
\mathop{U'_{3}}\limits^{\!\!\!k+3}+\cdot\cdot\cdot\right)\\+\left(k+\frac{1}{2}\right)
[\delta_{ik}-(g_3^++(-1)^kg_3^-)]=0,
\end{array}{}
\end{equation}
where $U'=(2k+1)\left(\mathop{U}\limits^{\!\!\!k+1}+\mathop{U}\limits^{\!\!\!k+3}+\cdot\cdot\cdot\right)$.
While systems (6.13) from [2] are true for $\forall N$.

Let us consider the problem of satisfaction of boundary conditions on $S^\pm$ for the class of
refined theories in the wide sense [2, Ch. I, 3]. Here is  necessary to note that, among
the refined theories we found that the models of Reissner and Ambartsumian satisfy these conditions.

By [3] we have
\begin{equation}
\begin{array}{l}
\displaystyle \sigma_{\alpha 3}=\frac{3Q_\alpha}{h}\left[1-\frac{z^2}{h^2}\right],\;\;\;
\sigma_{33}=-\frac{3q}{4}\left[\frac{2}{3}-\frac{z}{h}+\frac{1}{3}\left(\frac{z}{h}\right)^3\right],\\
\displaystyle \sigma_{33,3}|_{z=\pm h}=-\frac{3q}{4}\left[-\frac{1}{h}+\frac{z^2}{h^3}\right]_{z=\pm h}=0.
\end{array}{}
\end{equation}
Here for the linear case according to (2) $g_3^+=0,\;g_3^-=-q$.
We stress the fact that the boundary
condition (18): $\sigma_{33,3}|_{z=\pm h}=0$ is an artificial and odd condition.
Then the third
equation of (1) for the linear and isotropic case when $f_3\equiv0,\;-h\leq z\leq h$
is satisfied on $S^\pm$, i.e. $\sigma_{33,3}|_{z=\pm h}=0$.

In [4] is considered a geometrically classical nonlinear case when in (4):
$$
\displaystyle u_{k,i}u_{k,j}=u_{3,i}u_{3,j}\;(i,j=1,2),\;\;\;u_{k,3}u_{k,3-i}=0,\;(i=0,1,2),
$$
for a homogenous anisotropic plate with no more  than 13 independent constants
in Hooke's law (4) of elastic plates with constant thickness. Ambartsumian transferred
methodology of [3] satisfying the boundary conditions on the surfaces: $\sigma_{i3}|_{S^\pm}=g_i^\pm$ at
$\sigma_{33,3}^\pm=-\sigma_{13,1}^\pm-\sigma_{23,1}^\pm$.
If $f_3=g_\alpha^\pm=0$, $\sigma_{33,3}^\pm=0$ and he studied the general
case, $g_\alpha^\pm\neq0$, and expressed the tangential components of the stress vector as
$$
\displaystyle \sigma_{\alpha 3}=\frac{h+z}{2h}g_\alpha^++\frac{h-z}{2h}g_\alpha^-+\frac{4}{3h^3}(h^2-z^2)\varphi_{\alpha}(x,y),
$$
In this simple case from (1) foolows that: $\varphi_{\alpha,\alpha}=h(g_\alpha^++g_\alpha^-)+g_3^+-g_3^-$.

Then the nonlinear system of five DEs with respect to $u_\alpha(x,y)$, $w(x,y)$
(correspondingly of averaged values of horizontal and normal components of the displacement vector), $\varphi_\alpha$ is constructed.
The linear part of leading two DEs of this system are second order relatively to
$u_\alpha$, third one-the first order with respect to $\varphi_\alpha$,
the last two DEs have third order partial derivatives;
nonlinear part contain  correspondingly two, fourth and zero
degrees of product of the first and second orders derivatives under $w(x,y)$ The questions,
connected with these types of systems are open. Like in the linear case by $\sigma_{33}|_{z=\pm h}=g_3^\pm$ ,
according [3] $\sigma_{33,3}|_{z=\pm h}=\sigma_{\alpha3,\alpha}|_{z=\pm h}=g_{\alpha,\alpha}^\pm$
the same process was considered in the nonlinear case [4, ch. 2, §8], p.67: if $f_3=0,\;g_\alpha^\pm=0$, then
$$
\displaystyle \sigma_{33}=\frac{g_3^++g_3^-}{2}+\frac{3z}{4h^3}(g_3^+-g_3^-)\left(h^2-\frac{z^2}{3}\right),\;\;\Rightarrow
$$
$$
\displaystyle \sigma_{33}|_{S^\pm}=g_3^\pm,\;\;\;\sigma_{33,3}|_{S^\pm}=0,\;\;\;\sigma_{i3,i}(x,y,\pm h)=0.
$$

We remark that the complexity and obvious errors of the methodology in [4] are the result of
using of the expression of the first order of $\varphi_{\alpha,\alpha}$ in the systems of DEs almost everywhere
as well as the corresponding equation is the basis for constructing essential DE
with respect to an averaging deflection.  We underline  that the methodology
according  to [3-4] are popular and this approach were used by Timoshenko,
Donnel, Lukasiewicz, Morozov,$...$.

Below we try to use a careful and correct
approach to resolve this problem of satisfaction of (2) for sufficient
general cases. With this aim, we use a methodology from [2, point 6.3]
and consider the following relation for the nonlinear case too:
\begin{equation}
\begin{array}{l}
\displaystyle T_3(x,y,z)=\frac{(h+z)g_3^+}{2h}+\frac{(h-z)g_3^-}{2h}
\displaystyle +\mathop{\sum}\limits^{\infty}_{s=1}\mathop{T_s}\limits^{3} (x,y)
\left[p_{s+1}\left(\frac{z}{h}\right)-p_{s-1}\left(\frac{z}{h}\right)\right],
\\
\displaystyle T_{i3}|_{S^{\pm}}=\sigma_{i3}+\sigma_{j3}u_{i,j}=g_i^{\pm},\;\;\;x\in
S^{\pm}
\end{array}
\end{equation}

For the simplicity and clearness let us consider the case when Lam$\acute{\textmd{e}}$ coefficients
$\lambda,\;\mu$ are constants. In this case the uniform of refined theories corresponding to the
bending for deflection and generalized shearing forces is of the form
\begin{equation}
\begin{array}{l}
\displaystyle (D\Delta^2+2h\rho\partial_{tt}-2DE^{-1}(1+\nu)\rho\partial_{tt}\Delta)w=
\left(1-\frac{h^2(1+2\gamma)(2-\nu)}{3(1-\nu)\Delta}\right)\\
\displaystyle \times(g_3^+-g_3^-)+2h(1-\frac{2h^2(1+2\gamma)}{3(1-\nu)}\Delta)[w,\varphi]
+h(g_{\alpha,\alpha}^+-g_{\alpha,\alpha}^-)\\
\displaystyle -\int_{-h}^h\left(tf_{\alpha,\alpha}-\left(1-\frac{1}{1-\nu}\Delta(h^2-t^2)\right)f_3\right)dt,
\end{array}
\end{equation}
\begin{equation}
\begin{array}{l}
\displaystyle \left(\Delta^2-\frac{1-\nu^2}{E}\rho\Delta\partial_{tt}\right)\varphi=-\frac{E}{2}[w,w]\\
\displaystyle +\frac{\nu}{2}\left(\Delta-\frac{2\rho}{E}\partial_{tt}\right)(g_3^++g_3^-)+
\frac{1+\nu}{2h}f_{\alpha,\alpha}.
\end{array}
\end{equation}

Below we for simplicity consider only static cases. Then for an isotropic case we have:
\begin{equation}
\begin{array}{l}
\displaystyle D\Delta^2u^*_3=-\left(1-\frac{h^2(1+2\gamma)(2-\nu)}{3(1-\nu)}\Delta\right)\left(g_3^+-g_3^-\right)\\
\displaystyle +2h\left(1-\frac{2h^2(1+2\gamma)}{3(1-\nu)}\right)L\left[u^*_3,F_*\right]
+\left(g^+_{\alpha,\alpha}+g^-_{\alpha,\alpha}\right)\\
\displaystyle -\mathop{\int}\limits^{h}_{-h}\left(tf_{\alpha,\alpha}-
\left(1-\frac{1}{1-\nu}(h^2-t^2)\Delta\right)f_3\right)dt+R_8\left[u^*_3;\gamma\right],
\end{array}
\end{equation}
\begin{equation}
\begin{array}{l}
\displaystyle Q_{\alpha 3}-\frac{1+2\gamma}{3}h^2\Delta Q_{\alpha 3}=-D\Delta
u^*_3\\
\displaystyle +\frac{h^2(1+2\gamma)}{3(1-\nu)}\partial_\alpha\left(g_3^+-
g_3^-+2h(1+\nu)L\left[u^*_3,F_*\right]\right)\\
\displaystyle +\left(g^+_{\alpha}+g^-_{\alpha}\right)-
\mathop{\int}\limits^{h}_{-h}\left(tf_{\alpha}-
\frac{1+\nu}{2(1-\nu)}(h^2-t^2)f_{3,\alpha}\right)dt+R_{5+\alpha}\left[Q_{\alpha
3};\gamma\right],
\end{array}
\end{equation}

We underline that $L[u,v]=[u,v]=\partial_{11}u\partial_{22}v-2\partial_{12}u\partial_{12}v+\partial_{11}v\partial_{22}u$
is the well-known Monge-Amp$\acute{\textrm{e}}$re operator.
Now we investigate the influences of conditions (19) on the systems
of differential equations (22)-(23), construction of which
essentially depends on $Q_\alpha,\;\psi_\alpha,\;I=(T_{33},t)$,
$$
\displaystyle \psi_\alpha=\frac{1}{2}(h^2-t^2,\sigma_{\alpha 3}),
$$
$$
\displaystyle I=(T_{33},t),\;\;\;(u,v)=\int_{-h}^hu(x,y,t)v(x,y,t)dt.
$$

Evidently, we have:
$$
\displaystyle Q_{\alpha 3}=-2h\mathop{T_{\alpha
3}}\limits^{\!\!\!\!\!\!1}+h(g_\alpha^++g_\alpha^-),
$$
$$
\displaystyle \psi_\alpha=\frac{2h^3}{3}\left(g_\alpha^++g_\alpha^--\frac{17}{20}\mathop{T_{\alpha
3}}\limits^{\!\!\!\!\!\!1}\right)-\frac{2h^2}{3}\mathop{\int}\limits^{h}_{-h}
T_{\alpha 3}(x,y,t)p_2\left(\frac{t}{h}\right)dt
$$
$$
\displaystyle -\frac{1}{2}\mathop{\int}\limits^{h}_{-h}(h^2-t^2)\sigma_{i3}u_{\alpha,i}dt=
\frac{2h^3}{3}\left(g_\alpha^++g_\alpha^--\frac{17}{20}\mathop{T_{\alpha
3}}\limits^{\!\!\!\!\!\!1}\right)+R[\psi_\alpha],
$$
$$
\displaystyle \left|R[\psi_\alpha]\right|\leq h^{5/2}(c_1\|\sigma_{\alpha 3}\|+
c_2\|\sigma_{i3}u_{\alpha,i}\|),
$$
$$
\displaystyle I=\mathop{\int}\limits^{h}_{-h}tT_{33}dt=\frac{(1+2\gamma)h^2}{3}\left(g_3^+-g_3^-\right)
-\mathop{\int}\limits^{h}_{-h}t\sigma_{i3}u_{\alpha,i}dt,
$$
$$
\displaystyle \left\|\mathop{\int}\limits^{h}_{-h}t\sigma_{i3}u_{\alpha,i}dt\right\|=O(h^2).
$$

Now in (22)-(23) we change the components of normal rotations by functions
$\mathop{T_{\alpha 3}}\limits^{\!\!\!\!\!\!1}$. We can see that the boundary
condition on the surfaces satisfying all refined
theories depend on parameter $\gamma$. The system of DEs (22-23) contains Monge-
Amp$\acute{\textmd{e}}$re operator for the nonlinear case. As it is known, even in the case of an isotropic
elastic plate of constant thickness the subject of justification was an unsolved problem.
The point is that von K$\acute{\textmd{a}}$rm$\acute{\textmd{a}}$n, Love,
Timoshenko, L. Landau, Lukasiewicz, Washizu considered Saint-Venant-Beltrami
compatibility condition as one of the equations of the corresponding system
of DEs. In [11] we have proved that all DEs systems of von KMR type follow from (1).

We have the following relation (decomposition of Monge-Amp$\acute{\textmd{e}}$re operator):
$$
\displaystyle \partial_1[\partial_1(\partial_2u\partial_2v)-\partial_2(\partial_1u\partial_2v)]-
\partial_2[\partial_2(\partial_1u\partial_2v)-\partial_1(\partial_2u\partial_1v)]
$$
$$
\displaystyle =-(\partial_{11}u\partial_{22}v-2\partial_{12}u\partial_{12}v+\partial_{22}u\partial_{11}v).\;\;\;(M-A)
$$

It is necessary that to system (20)-(21) we must add, for evidence, part of von
K$\acute{\textmd{a}}$rm$\acute{\textmd{a}}$n type system (an isotropic case, see [11, formula (17)]:
\begin{equation}
\displaystyle (\lambda^*+2\mu)\partial_1\tau+\mu\partial_2\omega=\frac{1}{2h}\bar{f}_1
+\mu(\partial_1(\bar{u}_{3,2})^2-\partial_2(\bar{u}_{3,1}\bar{u}_{3,2}))+
\lambda_1(\sigma_{33,1},1),
\end{equation}
\begin{equation}
\displaystyle (\lambda^*+2\mu)\partial_2\tau-\mu\partial_1\omega=\frac{1}{2h}\bar{f}_2
+\mu(\partial_2(\bar{u}_{3,1})^2-\partial_1(\bar{u}_{3,1}\bar{u}_{3,2}))+
\lambda_1(\sigma_{33,2},1).
\end{equation}
Here $\displaystyle \tau=\bar{\varepsilon}_{\alpha\alpha},\;\omega=\bar{u}_{1,2}-\bar{u}_{2,1}$ are plane expansion and rotation,
$\displaystyle \lambda_1=\lambda/2h(\lambda+2\mu)$, nonlinear terms represent a
decomposition of Monge- Amp$\acute{\textrm{e}}$re operator if in $(M-A)$, $u=v=u_3$.

We must remark that in the general case one can find for the anisotropic case the
expressions (16) from [11]. The general transversality case if
$c_{11}=c_{22}=c_{12}+c_{66}$,
$b_{13}=b_{23}=b$, (see (22), [11]), might be interesting for applications in the following
form:
\begin{equation}
\begin{array}{l}
\displaystyle c_{11}\partial_1\tau+\frac{1}{2}c_{66}\partial_2\omega=\frac{1}{2h}\bar{f}_1\\
\displaystyle -bb_{33}^{-1}
\frac{1}{2h}\mathop{\int}\limits^{h}_{-h}\sigma_{33,1}dt-hc_{66}
[\partial_2(\bar{u}_{3,1}\bar{u}_{3,2})-\partial_1(\bar{u}_{3,2})^2]+R_1^A,
\end{array}{}
\end{equation}
\begin{equation}
\begin{array}{l}
\displaystyle c_{11}\partial_2\tau-\frac{1}{2}c_{66}\partial_1\omega=\frac{1}{2h}\bar{f}_2\\
\displaystyle -bb_{33}^{-1}
\frac{1}{2h}\mathop{\int}\limits^{h}_{-h}\sigma_{33,2}dt-hc_{66}
[\partial_1(\bar{u}_{3,1}\bar{u}_{3,2})-\partial_2(\bar{u}_{3,1})^2]+R_2^A.
\end{array}{}
\end{equation}

Below we give some extensions:

1. Let for anisotropic case the number of independent elastic
modulus are taken according to [2, 2.2]. If now we used the
corresponding expressions (2.17), (2.30) from [2]:
$$
\displaystyle u_\alpha^*=-u^*_{3,\alpha}+\frac{h^2(1+2\gamma)}{3D(1-\nu)}Q_{\alpha
3}+R_\alpha\left[u^*_\alpha;\gamma\right],
$$
$$
\displaystyle Q_{\alpha 3}=-\frac{2h^3}{3}L_{\alpha
3}(\partial_1,\partial_2)u_3^*+\frac{2h^3}{3}L_{\alpha\beta}(\partial_1,\partial_2)\psi_\beta^A+h(g_\alpha^++g_\alpha^-)
$$
$$
\displaystyle +\mathop{\int}\limits^{h}_{-h}t[(a\partial_1+b\partial_2)\sigma_{33}+cf_\alpha]dt+R_{2+\alpha}[Q_{\alpha
3};\gamma],
$$
where
$$
\psi_\alpha^A=3(\delta h^3)^{-1}\int_{-h}^htdt\int_0^t(b_{\alpha+3,\alpha+3}\sigma_{\alpha 3}-b_{45}\sigma_{33-\alpha})dt,
\;\;\;\delta=b_{44}b_{55}-b_{45}^2,
$$
$$
L_{\alpha 3}(\partial_1,\partial_2;c)=c_{\alpha\alpha}\partial_\alpha^3+3c_{\alpha 6}\partial_\alpha^2\partial_{3-\alpha}+
(c_{12}+2c_{66})\partial_\alpha\partial_{3-\alpha}^2+c_{3-\alpha6}\partial_{3-\alpha}^3,
$$
$$
L_{\alpha\alpha}(\partial_1,\partial_2;c)=c_{\alpha\alpha}\partial_\alpha^2+2c_{\alpha 6}\partial_\alpha\partial_{3-\alpha}+
c_{66}\partial_{3-\alpha}^2,
$$
$$
L_{12}=L_{21}=c_{16}\partial_1^2+(c_{12}+2c_{66})\partial_{12}+c_{26}\partial_2^2.
$$
As well as the equations (2.36-37) [2] with respect to
$u_3^*,\;Q_{\alpha 3}$ are the same to (22), (23), it would be
evident that all analogical conclusions are true also for the
anisotropic case.

2. We consider also the case when the Lam$\acute{\textrm{e}}$ coefficients are
variable. In this case, the Reissner type form for bending process
has the following face:
$$
\displaystyle \left[\left((1-\nu)D\mathop{\tau_{\alpha,1}}\limits^{\!\!\!\!\!\!\!\!\!\!1}\right)_{,1}+
\left((1-\nu)D\mathop{\tau_{\alpha,2}}\limits^{\!\!\!\!\!\!\!\!\!\!1}\right)_{,2}\right]
$$
$$
\displaystyle +\left[\left((1+\nu)D\mathop{\tau_{\alpha,\alpha}}\limits^{\!\!\!\!\!\!\!\!\!\!1}\right)_{,\alpha}+
\left(2\nu
D\mathop{\tau_{3-\alpha,3-\alpha}}\limits^{\!\!\!\!\!\!\!\!\!\!\!\!\!\!\!\!\!\!\!\!\!\!\!\!\!\!\!\!\!\!\!\!\!\!1}\right)_{,\alpha}+
\left((1-\nu)D\mathop{\tau_{3-\alpha,\alpha}}\limits^{\!\!\!\!\!\!\!\!\!\!\!\!\!\!\!\!\!\!\!\!\!\!\!\!\!1}\right)_{,3-\alpha}\right]
$$
$$
\displaystyle -\frac{3(1-\nu)}{h^2(1+2\gamma)}\left(\mathop{\tau_\alpha}\limits^{\!\!\!\!1}+u^*_{3,\alpha}\right)=
2(f_\alpha^*+R_{\alpha+2}),
$$
$$
\displaystyle \frac{3}{1+2\gamma}\left[\frac{1-\nu}{h^2}D\left(u^*_{3,\alpha}+\mathop{\tau_{\alpha}}\limits^{\!\!1}
\right)\right]=2\left(f_3^*+R_5\right).
$$

3. Let $\displaystyle \Omega_h=D(x,y)\times(h_1(x,y),h_2(x,y))$, $2h=h_2-h_1$, $\displaystyle \bar{h}=\frac{1}{2}(h_1+h_2)$.
For this case the function $\displaystyle v_\alpha(x,y,z)=u_{3,\alpha}-\frac{1}{\mu}\sigma_{\alpha 3}$
has important weight. Instead of (22)-(23) we have:
$$
\begin{array}{l}
\displaystyle \frac{1}{h^3}D\partial_\alpha\left(h^3\Delta u_{3,\alpha}\left(x,y,\bar{h}\right)\right)=
\sigma^+_{33}-\sigma^-_{33}\\
\displaystyle +\int_{h_1}^{h^2}f_3dt+\frac{h^2(2-\nu)(1+2\gamma)}{3(1-\nu)}(\Delta\sigma^+_{33}-\Delta\sigma^-_{33})+\Phi_1+R_1,\\
\displaystyle \frac{4h}{3}\sigma_{\alpha 3}(x,y,\bar{h})-\frac{2h^3}{3}\Delta\sigma_{\alpha 3}(x,y,\bar{h})\\
\displaystyle =-D\Delta u_{3,\alpha}(x,y,\bar{h})-\frac{1+2\gamma}{3(1-\nu)}(\sigma^+_{33,\alpha}-\sigma^-_{33,\alpha})+
\Phi_{1+\alpha}+R_{1+\beta}.
\end{array}{}
$$

We can see that for a variable thickness it  is possible to use the same methodology,
that was   used for the normal stress vector in such expressions:
$$
\begin{array}{l}
\displaystyle \sigma_3=\frac{h_2-z}{2h}g^-+\frac{z-h_1}{2h}g^++\mathop{\sum}\limits^{\infty}_{s=1}
\mathop{\tau_3}\limits^{s}(x,y)\left(p_{s+1}\left(\frac{2z-\bar{h}}{2h}\right)-p_{s-1}\left(\frac{2z-\bar{h}}{2h}\right)\right)
\\
\displaystyle =\left(a-\mathop{\tau}\limits^{1}\right)p_0+\left(b-\mathop{\tau}\limits^{2}\right)p_1+
\mathop{\sum}\limits^{\infty}_{s=2}\left(\mathop{\tau}\limits^{s-1}-\mathop{\tau}\limits^{s+1}\right)
p_{s}\left(\frac{2z-\bar{h}}{2h}\right),\;\;\;a,b=a,b(g^\pm,h),
\\
\displaystyle \sigma_3=\mathop{\sum}\limits^{\infty}_{s=0}\mathop{\sigma}\limits^{s}(x,y)
p_{s}\left(\frac{2z-\bar{h}}{2h}\right).
\end{array}{}
$$

4. New edge effect

If $T_{33}|_{S^\pm}=g_3^\pm$ then by the value
$$
\varphi=\int_{-h}^htT_{33}dt=\frac{(1+2\gamma)h^2}{3}(g_3^+-g_3^-)+R,\;\;\;(x,y\in\partial D).
$$
a new edge effect will be formed. This member is given in all boundary conditions
corresponding to (3) (for details see [2, ch. I, 3.3.1].

5. On applications of the complex variable function theory.

The representations (20)-(23) allow to apply complex analysis. Let us preliminarily
consider the equation (22) and underline the main members:
\begin{equation}
\begin{array}{l}
\displaystyle D'\Delta[u,\varphi]=D'([\Delta u,\varphi]+[u,\Delta\varphi]+2[\partial_\alpha u,\partial_\alpha\gamma]),\\
\displaystyle (D'=4h^3(1+2\gamma)/3(1-\nu)),\;D\Delta^2u.
\end{array}{}
\end{equation}
If $\{f\}$ denotes the physical dimension of value $f$ then it's evident that $\{\Delta^2u\}=\{\Delta[u,\varphi/E]\}$,
$E$ is  modulus of elasticity. We have:
\begin{equation}
\begin{array}{l}
\displaystyle [\Delta u,\varphi]=\partial_{22}\varphi\partial_{11}\Delta u-2\partial_{12}\varphi
\partial_{12}\Delta u+\partial_{11}\varphi\partial_{22}\Delta u,
\end{array}{}
\end{equation}
\begin{equation}
\begin{array}{l}
\displaystyle [\partial_\alpha u,\partial_\alpha \varphi]=\partial_{11\alpha}u\partial_{22\alpha}\varphi-
2\partial_{12\alpha}u\partial_{12\alpha}\varphi-\partial_{22\alpha}u\partial_{11\alpha}\varphi.
\end{array}{}
\end{equation}
Thus, the first summand type (29) of (28) may be define also the nonlinear wave
processes in the static cases whereas the third order derivatives containing a summand
(30) with respect to function $u=u(x,y)$ corresponds to 1 and 2-dim soliton  solutions
of Corteveg-de Vries or Kadomtsev-Petviashvili kind. As the second order derivatives
of  the function $\varphi=\varphi(x,y)$ describe  the stress tensors horizontal components, the summands
$[u,\Delta\varphi]$ correspond to the nonlinear part for the systems of the type (22-23).

The calculation and analysis of a symbolical determinant of these expressions show
that the characteristic forms of the systems of type (22-23) may be positive, negative or
zero  as they represent arbitrary functions of $x,\;y$. Let us consider the following
operators and notations:
$$
z=x+iy,\;\;\;\bar{z}=x-iy,\;\;\;x_1=x,\;x_2=y,\;u(x,y)=U(z,\bar{z}),
$$
$$
\displaystyle \partial_{\bar{z}}=\frac{1}{2}(\partial_1+i\partial_2),\;\;\;\partial_{z}=\frac{1}{2}(\partial_1-i\partial_2),
$$
$$
4\partial_{\bar{z}z}=\Delta,\;\;\;16\partial_{\bar{z}z}\partial_{\bar{z}z}=\Delta^2,\;\;\;
16\partial_{\bar{z}z}[U(\bar{z},z),V(\bar{z})]=\Delta[u(x,y),v(x,y)].
$$
Now we form the following iterative-direct (hybrid) method for finding
the solution of rewriting in complex variables  systems of PDEs (23)-(26) so:

Let $[U(z,\bar{z})]^{[m]}$ denotes $m-th$ approach for deflection $u_3^*(x,y)$ which is
calculated by known right-hand terms without $R$ and $m-1$-th order approach of summand
\begin{equation}
\begin{array}{l}
\displaystyle \frac{2Eh}{16D}\left(1-\frac{h^2(1+2\gamma)}{3(1-\nu)}\partial_{\bar{z}}
\partial_z\right)\int_0^{z}\int_0^{\bar{z}}(z-\zeta)(\bar{z}-\bar{\zeta})[U,V]^{[m-1]}d\zeta d\bar{\zeta},\\
\displaystyle EV=\Phi\left(\frac{z+\bar{z}}{2},\frac{z-\bar{z}}{2i}\right),
\end{array}{}
\end{equation}
We do some operations for DEs (23) for shearing forces and for system (24-25). This
system is equivalent to the following equation (see [11]):
$$
\displaystyle \Delta(\sigma_{11}+\sigma_{22})=-\frac{E}{2}[w,w]+\frac{\nu}{2h}
\int_{-h}^h\Delta\sigma_{33}dt+\frac{1+\nu}{2h}\bar{f}_{\alpha,\alpha},\;\;\;(K-R)_2.
$$
For
\begin{equation}
\begin{array}{l}
\displaystyle V^{[m]}=V^{[m]}(z,\bar{z})=-\frac{\mu}{\lambda^*+2\mu}
\int_0^{\bar{z}}\int_0^{z}(\bar{z}-\bar{\zeta})(z-\zeta)[U^{[m-1]},U^{[m-1]}]d\vec{\zeta}d\zeta+F(\bar{z},z).
\end{array}{}
\end{equation}

We remark :

i. The correction of $(K-R)_2$ equation by summand depending
from $\Delta\sigma_{33}$ was considered by Lukasiewicz considering
only effects of local loads [12],

ii. The invariant form of Monge- Amp$\acute{\textmd{e}}$re operator:
$$
\displaystyle [u(x,y),v(x,y)]=-4[\partial_{\bar{z}\bar{z}}U\partial_{zz}V-2\partial{\bar{z}z}U\partial_{\bar{z}z}V+
\partial_{zz}U\partial_{\bar{z}\bar{z}}V]
$$
$$
\displaystyle =-4[U(z,\bar{z}),V(z,\bar{z})],\;\;\;(M-A)_{cf},
$$
and when $u=v$ we have
$$
\displaystyle [u,u]=2\left(\partial_{11}u\partial_{22}u-\left(\partial_{12}u\right)^2\right)=-4[U(z,\bar{z}),V(z,\bar{z})].
$$
Thus, by means of complex analysis we reduced the systems of PDEs of KMR type to
the pseudo-integral operator of second type. An iterative scheme, described by
(31) corresponds to the solution of Volterra second type nonlinear
integral equation. Whereas the processes by schemes generating from
(22) contain  both Volterra and Fredholm type operators with an
arbitrary parameter $\gamma$. The convergence  for only pure Volterra
type process (where $\gamma=-0.5$) depends also on the convenient selection of
the initial functions $U^{[0]}$, $V^{[0]}$. It is possible to apply some results
of [13, Ch. XXIV, §476, example 4] to the equation $[u,u]+a^2=0$ for arbitrary
function $a=a(x,y)$. When $\gamma\neq0.5$ the convergence depends on the Fredholm operator:
$$
F_r(U,V)=\partial_{\bar{z}}\partial_z\lambda\int_0^{\bar{z}}\int_0^z(\bar{z}-\bar{\zeta})(z-\zeta)
[U(\zeta,\bar{\zeta})V(\zeta,\bar{\zeta})]d\bar{\zeta}d\zeta
$$
with an arbitrary parameter denoted for simplicity by $\lambda$. The operator $\lambda^{-1}F(U,V)$
depends on the behavior of expression which may generate different kinds of waves (shok, soliton)
functions too and in the cases when they are uniformly bounded functions the process
corresponding to applications of the Fredholm operator will be
convergent as the corresponding operator will be a contracted one.
More convenient may be Seidel's type iterative scheme: let the initial value is $\displaystyle U^{[0]}=\frac{1}{4}z^2\bar{z}^2$.
Then in expressions of type (31) we used $V^{[1]}$ defining from (32) and so on. The following
theorem is true

\textbf{Theorem 4.} \textit{Let us consider the following iterative process:}
$$
\begin{array}{l}
\displaystyle V^{[m]}(z,\bar{z})=a\int_0^z\int_0^{\bar{z}}(z-\zeta)(\bar{z}-\bar{\zeta})
\left[U^{[m-1]},U^{[m-1]}\right]d\zeta d\bar{\zeta},\;\;\;m=1,2,...,\\
\displaystyle U^{[m]}(z,\bar{z})=b\int_0^z\int_0^{\bar{z}}(z-\zeta)(\bar{z}-\bar{\zeta})
\left[U^{[m-1]},V^{[m]}\right]d\zeta d\bar{\zeta}\\
\displaystyle +c\int_0^z\int_0^{\bar{z}}\left[U^{[m-1]},V^{[m]}\right]d\zeta d\bar{\zeta},\;\;\;m=1,2,...,
\end{array}{}
$$
\textit{then it is convergence for all finite $\displaystyle a,\;b,\;|c|<\frac{4}{3}$,
$U^{[0]}=z^n\bar{z}^n$ and an integer $\forall n\geq2$.}

\textit{Proof.} The essential moment is to estimation of the
transition effect from $m$ step to
$m+1$ step. Let $U^{[m]}=z^p\bar{z}^p$. The transition process contains two stages: the
calculation of  expressions of the  type $[u,v]$ and corresponding integrals. It is evident that
$$
\displaystyle \left[U^{[m]},U^{[m]}\right]=2(p(p-1))^2z^{2p-2}z^{2p-2}-2p^4z^{2p-2}\bar{z}^{2p-2}=
-2p^2(2p-1)z^{2p-2}\bar{z}^{2p-2},
$$
then we have also:
$$
\displaystyle V^{[m+1]}=-\frac{2ap^2(2p-1)}{4p^2(2p-1)^2}z^{2p-2}\bar{z}^{2p-2}=-\frac{a}{2(2p-1)}z^{2p-2}\bar{z}^{2p-2},
$$
and
$$
\displaystyle \left[U^{[m]},V^{[m+1]}\right]=-\frac{2ap^2(3p-1)}{2p-1}z^{2p-2}\bar{z}^{2p-2},\;\;\;c_p=\frac{2p^2(3p-1)}{2p-1},
$$
$$
\displaystyle I_1=abc_p\int_0^z\int_0^{\bar{z}}z^{2p-2}\bar{z}^{2p-2}dzd\bar{z}=\frac{abc_p}{4p^2(2p-1)^2}z^{2p}\bar{z}^{2p},
$$
$$
\displaystyle I_2=acc_p\int_0^z\int_0^{\bar{z}}z^{2p-2}\bar{z}^{2p-2}dzd\bar{z}=\frac{acc_p}{(2p-1)^2}z^{2p-1}\bar{z}^{2p-1},
$$
$$
\displaystyle c_p(2p-1)^{-2}<\frac{3}{4}+\frac{7}{8(p-1,5)}.
$$
This relation show that if $\displaystyle \left|c=\gamma\right|<\frac{4}{3}$ for all bounded functions $a,b$
the above iterative process is convergence.

Remark.  We calculated the  systems (22), (23)
approximately by the Euler-McLaurin quadrature formulae
the summands of members contain $(\Delta\sigma_{33},1)$ by $\Delta\sigma_{33}(x,y,\pm h)$ and
$\lim\Delta\sigma_{33,3}(x,y,\pm h-0)$. Then we
use the explicit representation of the Cauchy-Riemann
nonhomogeneous system of DEs with respect to
$$
\displaystyle w(z,\bar{z})=(\lambda^*+2\mu)\tau\left(\frac{z+\bar{z}}{2},\frac{z-\bar{z}}{2i}\right)+
i\mu\omega\left(\frac{z+\bar{z}}{2},\frac{z-\bar{z}}{2i}\right),
$$
for
$$
\partial_{\bar{z}}w(z,\bar{z})=F(z,\bar{z}),
$$
by Pompeiu formula (see i.e. (4.11) or (4.13) [14, ch. I, 4]).

The same processes are true for the anisotropic cases.

6. It is possible to use for an approximate solution by
numerical methods the systems of type (24-25), (26-27)
with boundary conditions generating by (3).

\section{Conclusion}.

Thus, we created the mathematical theory for refined theories both in linear and
nonlinear cases for anisotropic nonhomogeneous elastic plates and shells, approximately
satisfying the corresponding system of partial differential equations and
boundary conditions on the surfaces. Now, the optimal and convenient refined theory
might be chosen easily by selection of  the parameter $\gamma$  after making a few necessary
experimental measurements without using any simplifying hypotheses. We justified
and give the right form to the von K$\acute{\textmd{a}}$rm$\acute{\textmd{a}}$n-Reissner-Mindlin type systems of refined
theories. We demonstrated that the Monge-Amp$\acute{\textmd{e}}$re operator is a linear differential form of
the first order of two nonlinear operators having  applications in the nonlinear
elasticity theory and has invariant form within to sign  transform  from the real to the complex variables.

\begin{acknowledgement}
This research was supported by helping of Tsitsino Gabeskiria, Bakur Gulua, David Natroshvili, Nona Vasilieva--Vashakmadze.
\end{acknowledgement}

%%%%%%%%%%%%%%%%%%%%%%%% referenc.tex %%%%%%%%%%%%%%%%%%%%%%%%%%%%%%
% sample references
% %
% Use this file as a template for your own input.
%
%%%%%%%%%%%%%%%%%%%%%%%% Springer-Verlag %%%%%%%%%%%%%%%%%%%%%%%%%%
%
% BibTeX users please use
% \bibliographystyle{}
% \bibliography{}
%
\biblstarthook{
}

\end{document}